\documentclass[12pt]{article}
\usepackage{amssymb}

\usepackage{color}

\def\X{\mathcal X}
\def\P{\mathcal P}

\def\V{\mathcal V}
\def\T{\mathcal T}
\def\pd#1#2{\frac{\partial#1}{\partial#2}}
\def\DD<#1>{\left\langle\!\!\!\left\langle#1\right\rangle\!\!\!\right\rangle}
\def\dd<#1>{\langle\!\langle#1\rangle\!\rangle}
\def\<#1>{\langle#1\rangle}

\newtheorem{theorem}{Theorem}
\newtheorem{example}{Example}
\begin{document}

\markboth{J.F.\ Cari\~nena, I.\ Gheorghiu, E.\ Mart\'{\i}nez, and P.\ Santos}
{Virial theorem in quasi-coordinates and Lie algebroid formalism\ }
%

\title{\uppercase{Virial theorem in quasi-coordinates and Lie algebroid formalism}}

\author{%
{Jos\'{e} F. Cari\~{n}ena}$^a$,
{Irina Gheorghiu}$^b$,\\
{Eduardo Mart\'{\i}nez}$^c$,  and
{Patr\'{\i}cia Santos}$^{d}$\\[4pt]
$^a$ IUMA and Department of Theoretical Physics, \\ University of Zaragoza, \\
{jfc@unizar.es}\\
$^b$ Department of Theoretical Physics,\\ University of Zaragoza, \\
{irinagh@unizar.es}\\
$^c$ IUMA and Department of Applied Mathematics,\\ University of Zaragoza, 	\\	
{emf@unizar.es}\\
$^d$ CMUC, University of Coimbra, and \\ Polytechnic Institute of Coimbra, ISEC, Portugal\\ 
{patricia@isec.pt}%
}

\maketitle
\date{}

\begin{abstract}
In this paper, the geometric approach to the virial theorem developed  in \cite{CFR12} is written in terms of quasi-velocities (see \cite{CNCS07}). A generalization of the virial theorem for mechanical systems on Lie algebroids is also given, using the geometric tools of Lagrangian and Hamiltonian mechanics on the prolongation of the Lie algebroid.
\end{abstract}

\section{Introduction}
The virial theorem was introduced by Clausius  in statistical mechanics in 1870 and since then it became important in many other areas in physics (see \cite{GC78} for an historical account). In the original formulation the virial theorem establishes a relationship between the time averages of the kinetic energy and of the scalar product $\mathbf{r} \cdot \mathbf{F}$ of trajectory by force. In the particular case of a conservative system with homogeneous potential, this amounts to a relation between time averages of the kinetic and the potential energy. The modern approach to the virial theorem given in \cite{CFR12} uses Hamiltonian formalism and establishes that, under the general conditions of application of the theorem, the time average of the Poisson bracket of an observable $G$ with the Hamiltonian vanishes, obtaining as a particular case that of a regular Lagrangian system. The original statement of the virial theorem is recovered when $H$ is a sum of a kinetic and a potential term, $H=T+V$, and the virial function is $G=\mathbf{r}\cdot \mathbf{p}$, so that the Hamiltonian vector field $X_G$ is precisely the generator of dilations in phase space. But virial-like theorems are also available for systems with configuration space different from $\mathbb{R}^N$ \cite{LZC11}.

More concretely, the geometric version of the Virial Theorem (VT) given in~\cite{CFR12} is as follows. On a Poisson manifold $(M,\{\cdot,\cdot\})$, every function $H\in C^\infty(M)$ defines a dynamical system by $\dot{x}=X_H(x)=\{x,H\}$. Then, if a function $G$ remains bounded, the time average of the Poisson bracket $\{G,H\}$ vanishes:
\[
\dd<\{G,H\}>=0.
\]
By the time average we mean $\dd<F>=\lim_{T\to\infty}\frac{1}{T}\int_0^T F(\gamma(t))\,dt$, where $\gamma$ is the evolution curve.
In what follows we  implicitly assume that the virial function $G$ remains bounded. When
the motion of the dynamical system is periodic with period $\tau$ the time average reduces to $\dd<F>=\frac{1}{\tau}\int_0^\tau F(\gamma(t))\,dt$.


\section{Virial theorem in quasi-coordinates}

In many problems in classical mechanics and control theory it is useful to consider quasi-velocities.
For instance, in studying the rotation of a rigid body, it is traditional to use Euler's angles to parametrize the orientation of the body while using body angular velocities to describe the dynamics. Similarly, for a system with nonholonomic constraints (i.e. constraints on the velocities that are not derivable from position constraints) one can define quasi-velocities in such a way that some of them coincide with the constraints, obtaining in this way fewer equations to solve.
\subsection{Quasi-velocities and quasi-momenta}
We consider a configuration manifold $Q$ where a mechanical system is evolving.
The traditional concept of velocities and momenta are obtained when considering a local chart $(U,q^1,\ldots,q^n)$, and taking the coordinate basis $\{\partial/\partial q^j\}$ and its dual $\{dq^j\}$. Then, $v=v^j\partial/\partial q^j$ and $\zeta=p_j\,dq^j$, with  $v^j=\<dq^j,v>$ and $p_j=\<\zeta,\partial/\partial q^j>$ being the usual velocities and momenta.
Alternatively we can chose a local basis of vector fields on $Q$,
$\{X_1,\ldots,X_n\}$,  and  the dual basis $\{\alpha^1,\ldots,\alpha^n\}$. Any tangent vector $v\in T_qQ$ can be expressed uniquely as $v=w^jX_j(q)$. The real numbers $(w^1,\ldots,w^n)$ are called the \textit{quasi-velocities} of $v$ in the given basis. In terms of the dual basis $w^j=\langle\alpha^j(q),v\rangle$. Similarly a covector $\zeta\in T^*_qQ$ can be expressed as $\zeta=\pi_j\alpha^j(q)$, and then $(\pi_1,\ldots,\pi_n)$ are called the quasi-momenta of $\zeta$ in the given basis, which can be obtained as $\pi_j=\langle{\zeta},X_j(q)\rangle$.
The pair $(q^i,w^i)$ is called the quasi-coordinates of $v\in TQ$ and the pair $(q^i,\pi_k)$ is called the quasi-coordinates of $\zeta\in T^*Q$ (see \cite{CNCS07}).

The relation between standard velocities and quasi-velocities is given by the well-known basis change formulas. If $X_j=\beta_{j}^{k}(q)\partial_{q^k}$ is the coordinate expression of the vector field $X_j$ in the coordinate basis then $dq^j=\beta^j_k(q)\alpha^k$ and it follows that
$v^i=w^j\beta_{j}^{i}(q)$ and $\pi_k=p_i\beta_{k}^{i}(q)$.
A system of quasi-coordinates has an associated set of local functions on $Q$ called  Hamel's symbols given by
$
\gamma_{ml}^k=\beta^{j}_{m}\beta^{i}_{l}\bigl(\pd{\alpha^k_{j}}{q^i}-\pd{\alpha^k_{i}}{q^j}\bigr)
$, where $[\alpha^{i}_{m}]$ is the inverse matrix of $[\beta^{m}_{j}]$, i.e.\ $\alpha^{i}_{m}\beta^{m}_{j}=\delta^i_{j}$.  They can be defined by means of $d\alpha^k=-\frac{1}{2}\gamma^k_{ml}\alpha^m\wedge\alpha^l$, or alternatively by $[X_m,X_l]=\gamma^k_{ml}X_k$.


\subsection{VT in the Hamiltonian formalism and quasi-momenta}

A function in the phase space, $H\in C^\infty(T^*Q)$, determines  an associated Hamiltonian vector field $X_H$ by the dynamical equation $i_{X_H}\omega_0=dH$, where $\omega_0=-d\theta_0$ is the canonical symplectic form on $T^*Q$. The motions of the system are the integral curves of $X_H$. In quasi-coordinates $(q^i,\pi_i)$ on the cotangent bundle $T^*Q$, the differential of an arbitrary function $G\in C^{\infty}(T^*Q)$ is given by
$
dG=X_j(G)\alpha^j+\pd{G}{\pi_j}d\pi^j
$.
The canonical 1-form $\theta_0$ has the expression $\theta_0=\pi_k\alpha^k$ and the canonical symplectic form $\omega_0=-d\theta_0$ is locally given by
$
\omega_0=\alpha^i\wedge d\pi_i+\frac{1}{2}\pi_k\gamma^k_{ij}\alpha^i\wedge\alpha^j
$.
Therefore, the Hamiltonian vector field associated to the function $G$ is given by
\[
X_G=\pd{G}{\pi_i}X_i-\Bigl(\beta_i^j\pd{G}{q^j}+\pi_k\gamma^k_{ij}\pd{G}{\pi_j}\Bigr)\pd{}{\pi_i}.
\]

Given a virial function $G$ on $T^*Q$,  the virial theorem in the Hamiltonian formulation written in quasi-coordinates is
\[
\DD<\beta^j_i\pd{G}{\pi_i}\pd{H}{q^j}-\beta^j_i\pd{G}{q^j}\pd{H}{\pi_i}-\pi_k\gamma^k_{ij}\pd{G}{\pi_j}\pd{H}{\pi_i}>=0.
\]
This equation provides a geometric interpretation of the  VT as presented
in \cite{QG98} by using the Poincar\'{e}'s formalism. Particularly important are fibrewise linear virial functions. Every vector field $D$ on the base manifold $Q$ is associated with a linear function $G\in C^\infty(T^*Q)$ defined by $G(\zeta)=\langle\zeta,D(q)\rangle$ for $\zeta\in T^*_qQ$. The associated Hamiltonian vector field is the complete lift $D^c$ of $D$ to $T^*Q$. In quasi-coordinates $(q^i,\pi_i)$ on $T^*Q$, if $D$ has the expression $D=f^iX_i$ then $G(q,\pi)=\pi_kf^k(q)$ and the Hamiltonian vector field has the expression
\[
X_G=D^c=f^iX_i-\Bigl(\beta_i^j\pd{f^k}{q^j}+\gamma^k_{ij}f^j\Bigr)\pi_k\pd{}{\pi_i}.
\]
For such a function the virial theorem can be expressed in the form
\begin{equation}\label{VTLFqc}
\DD<\beta^j_if^i\pd{H}{q^j}-\beta^j_i\pd{f^k}{q^j}\pi_k\pd{H}{\pi_i}-\pi_k\gamma^k_{ij}f^j\pd{H}{\pi_i}>=0.
\end{equation}


\subsection{VT in the Lagrangian formalism and quasi-velocities}
Consider now a dynamical system defined by a regular Lagrangian $L\in C^\infty(TQ)$. The dynamical vector field $\Gamma_L\in\mathfrak{X}(TQ)$ is determined by the dynamical equation $i_{\Gamma_L}\omega_L=dE_L$, where $\omega_L=-d\theta_L$ is the Cartan 2-form associated to the Lagrangian and $E_L$ is the energy function defined by $L$. In quasi-coordinates $(q^i,w^i)$ on the tangent bundle $TQ$, the differential of an arbitrary function $G\in C^{\infty}(TQ)$ is given by
$dG=X_j(G)\alpha^j+\pd{G}{w^j}dw^j$,
and the Cartan 2-form $\omega_L=-d\theta_L$ by
$$
\omega_L=\frac{1}{2}\left[\gamma^k_{ml}\pd L{w^k}%
+X_l\left(\pd{L}{w^m}\right)-X_m\left(\pd{L}{w^l}\right)\right]
\alpha^m\land \alpha^l +\pd{^2L}{w^j\partial w^k}\,%
\alpha^k\land dw^j\ .
$$
Therefore, the dynamical vector field $\Gamma_L=X_{E_L}$ is given by
$$
\Gamma_{L}= w^jX_j+
  W^{r l}\left[w^m\gamma^k_{m l}\frac{\partial L}{\partial w^k}
  -w^m X_m\left(\frac{\partial L}{\partial w^l}\right)+
  X_l(L)\right]\pd{}{w^r},
$$
where $[W^{rl}]$ is the inverse matrix of $[\partial^2 L/\partial w^l\partial w^r]$, and the Hamiltonian vector field of the function $G$ is
$$
X_{G}= W^{jl}\pd{G}{w^l} X_j
      +W^{jl}\left\{\left[\pd{L}{w^k}\gamma^k_{m l}+X_l\left(\pd{L}{w^m}\right) -X_m\left(\pd{L}{w^l}\right)\right]
       W^{m r}\pd{G}{w^r}
      - X_l(G)\right\} \pd{}{w^j}.
$$

For a virial function $G$ the virial theorem takes the form
\begin{equation}
\DD<  \pd{G}{w^r} W^{r l}\left[w^m X_m\left(\frac{\partial L}{\partial w^l}\right) - X_l(L) - w^m\gamma^k_{m l}\frac{\partial L}{\partial w^k} \right]-w^jX_j(G)>=0
\end{equation}
The above equation provides a geometric interpretation of the Boltzmann's formalism of the VT. An important case is that of the function $G=\langle\theta_L,D^c\rangle$, where $D$ is a vector field on $Q$ and $D^c$ is its complete lift to $TQ$. It was proved in~\cite{CFR12} that $\{G,E_L\}=\Gamma_LG=D^cL$, from where it follows that on the condition of the virial theorem we have $\dd<D^cL>=0$. In quasi-coordinates, if $D=f^iX_i$ then the expression of the complete lift is
\[
D^c= f^iX_i +\left[X_k(f^i)+\gamma^i_{kj}f^j\right]w^k\pd{}{w^i},
\]
and therefore
\[
\DD<
f^iX_i(L) +\left[X_k(f^i)+\gamma^i_{kj}f^j\right]w^k\pd{L}{w^i}
>=0.
\]
If moreover the Lagrangian is of mechanical type, $L=T-V$, then the VT has the form $\dd<D^c(T)>=\dd<D(V)>$. In coordinates, turns out to be
\begin{equation}\label{VTLFqcM}
\DD<  f^iX_i(T)+\left[X_k(f^i)+\gamma^i_{kj}f^j\right]w^k\pd{T}{w^i}>=\dd<f^jX_j(V)>.
\end{equation}

\begin{example}[Kepler problem and quasi-velocities]
Let us consider a particle $P$ of mass $m$ moving in a plane under the action of a central force $F(r)=-\gamma mm'/r^2$ on the direction of a fixed point $O$ of mass $m'\gg m$, where $\gamma$ is a positive constant and $r$ represents the distance between $O$ and the particle $P$. The configuration space of the system is $Q=\mathbb{R}^2-\{O\}$. Let $\theta$ be the angle that the line $OP$ makes with a fixed direction on the plane. Consider as quasi-velocities $w^1=\dot r$ and $w^2=r^2 \, \dot \theta$, corresponding to twice the area swept-out per time unit. Then,
$$
\mathbf{L}(r,\theta,w^1,w^2)= \frac{m}{2}\left[(w^1)^2+\frac{1}{r^2}(w^2)^2\right]+\frac{\gamma mm'}{r}.
$$
Let $D=r\partial_r$ be the infinitesimal generator of dilations on the space $\mathbb{R}^2$ written in polar coordinates. The complete lift of $D$ is the vector field $D^c=r\partial_r+w^1\partial_{w^1}+2w^2\partial_{w^2}$ on the tangent bundle $T\mathbb{R}^2$. If the virial function is defined by $G=\langle\theta_L,D^c\rangle$, that is, $G(r,\theta,w^1,w^2)=m rw^1$, then the Hamiltonian vector field of $G$ turns out to be $X_G=r\partial_r-w^1\partial_{w^1}$. Applying formula (\ref{VTLFqcM}), we obtain
$\dd<r\partial_r V> = \dd< m(w^1)^2+m\frac{(w^2)^2}{r^2}>$,
that is, $\dd<-V> = \dd<2T>$ as expected.
\end{example}


\section{Virial theorem for mechanical systems on Lie algebroids}
The geometrical interpretation for quasi-coordinates has been given in \cite{CNCS07} which amounts to forget the tangent structure and to use only the vector bundle structure $\pi:TQ\to Q$ and the Lie algebra structure $[\cdot,\cdot]$ on the set of vector fields on $Q$. This more general framework leads naturally to the use of Lie algebroids. In this formalism we do not have a prefered basis of sections in the bundle induced by the choice  of coordinates in the base.

A Lie algebroid is a vector bundle $\tau:A\to M$, together with a Lie algebra structure $[\cdot ,\cdot ]$ in the space of its sections, and a vector bundle map $\rho :A\to TM$ over the identity in the base $M$, called the anchor, that satisfies the Leibniz rule $[\sigma,f \, \eta]=f\,[\sigma,\eta] +(\rho(\sigma)f)\eta,$ for any pair of sections $\sigma,\eta$ of $\tau:A\to M$ and every smooth function $f\in C^\infty(M)$. In the above expression, by $\rho(\sigma)$ we mean the vector field in $M$ given by $\rho(\sigma)(q)=\rho(\sigma(q))$ for every $q\in M$. In this way every section of $A$ defines a dynamical system $\dot{x}=\rho(\sigma)(x)$ on the base manifold $M$, whose orbits are the integral curves of the vector field $\rho(\sigma)$.

The Lie algebroid structure is equivalent to the existence of an exterior differential operator $d:\Omega^\bullet(A)\to \Omega^{\bullet+1}(A)$ on the exterior algebra of $A$-forms, i.e. the algebra of sections of the exterior product $\bigwedge^\bullet A^*\to M$, similar to the de Rham operator on a manifold and also satisfying $d^2=0$.

A choice of local coordinates $(x^1,\ldots,x^n)$ on the base manifold $M$ and a choice of a local basis of sections $\{e_\alpha\}$  of $A$ determines a local coordinate system $(x^i, y^\alpha)$ on the Lie algebroid $A$. The anchor and the bracket are locally determined by functions $\rho^i_{\,j}$ and $C^\gamma_{\alpha\beta}$ on $M$, called structure functions, determined by
$
\rho(e_\alpha)=\rho^i_{\alpha}\pd{}{x^i}\quad\mbox{and}\quad [e_\alpha,e_\beta]=C^\gamma_{\alpha\beta}\,e_\gamma
$.
The structure functions are not arbitrary but satisfy a set of structure equations, that are the local equivalent of the Leibniz identity and the Jacobi identity of the bracket.
\subsection{Symplectic Lie algebroids and general Hamiltonian systems}

By a symplectic Lie algebroid we mean a pair $(A,\omega)$ where $A$ is a Lie algebroid and $\omega\in\Omega^2(A)$ is a symplectic $2$-section, i.e. regular as a bilinear form and closed with respect to the exterior differential operator $d$ of the Lie algebroid $A$ (see \cite{LMM05} for the details). On a symplectic Lie algebroid $(A,\omega)$ every function $H\in C^\infty(M)$ defines a dynamical system on the base manifold $M$ as follows. Given the function $H$, there is a unique section $\sigma_H$ of  $A$, called Hamiltonian section of $H$, such that $i_{\sigma_H}\omega=dH$. The vector field $X_H=\rho(\sigma_H)$ is the infinitesimal generator of such a dynamical system.

The Hamiltonian vector field $X_H$ can also be obtained in terms of a Poisson bracket on the base manifold $M$. Indeed, given two function $F,G\in C^\infty(M)$, the bracket defined by $\{F,G\}=\omega(\sigma_F,\sigma_G)$ is a Poisson bracket on $M$. We clearly see the relations $\{F,G\}=i_{\sigma_G}dF=\rho(\sigma_G)F=X_GF=-X_FG$.

For the Poisson structure defined by the symplectic section $\omega$ on the Lie algebroid $A$, the VT implies that $\dd<\rho(\sigma_G)H>=0$. In coordinates, the differential of $G\in C^\infty(M)$ is given by $dG=\rho^i_\alpha\frac{\partial G}{\partial x^i}e^\alpha$, where $\{e^\alpha\}$ is the dual basis of sections of $A^*$. If the symplectic form is locally $\omega=\omega_{\alpha\beta}e^\alpha\wedge e^\beta$ then the VT can be written in the following way
\begin{equation}\label{VTLAF-c}
\DD<\omega^{\alpha\beta}\rho^i_\alpha\rho^j_\beta\pd{H}{x^i}\pd{G}{x^j}>=0,
\end{equation}
where $[\omega^{\alpha\beta}]$ is the inverse matrix of $[\omega_{\alpha\beta}]=[\omega(e_\alpha,e_\beta)]$.

The above construction depends on the availability of a symplectic section on the Lie algebroid $A$. In the next section we will show two important classes of symplectic Lie algebroids which generalize the standard Lagrangian and Hamiltonian approaches of the classical mechanics.

\subsection{Lagrangian and Hamiltonian systems on Lie algebroids}

The Lie algebroid approach to Hamiltonian and Lagrangian mechanics builds on the geometrical structure of the $A$-tangent to a fibre bundle $P$ over the same base, also called the prolongation of $P$ with respect to the Lie algebroid $A$ (see \cite{LMM05,EM01}).

Consider a Lie algebroid $\tau:A\to M$ and let $\nu:P\to M$ be a fibre bundle over the same base manifold $M$. For each point $p\in P$ we consider the vector space
$\T^A_{p}P =\{(a,v)\in A_q\times T_pP\mid \rho(a)=T_p\nu(v)\}$, where $q=\nu(p)$. The element $(a,v)\in\mathcal{T}_{p}P$ will be denoted by $(p,a,v)$. The manifold $\T^AP=\cup_{p\in P}\T^AP$ has a natural vector bundle structure over $P$ with the projection $\nu_1:\T^A P\to P$ given by $\nu_1(p,a,v)=p$. Moreover, it can be endowed with a natural Lie algebroid structure, as follows. The anchor map $\varrho:{\T^AP}\to TP$ is the projection onto the third factor and the bracket $[\![\cdot,\cdot]\!]$ is determined by imposing that the bracket of two projectable sections $\mathcal{Z}_i(p)=(p,\sigma_i(\nu(p)), U_i(p))$, $i=1,2$, is $[\![\mathcal{Z}_1,\mathcal{Z}_2]\!](p)=(p,[\sigma_1,\sigma_2](\nu(p)),[U_1,U_2](p))$. Local coordinates $(x^i,u^J)$ on $P$ and a local basis $\{e_\alpha\}$ of sections of $A$ determine a local basis of projectable sections of $\T^AP$ by $\mathcal{X}_\alpha(p)=(p,e_\alpha(\nu(p)),\rho^i_{\,\alpha}\partial_{x^i}|_p)$ and $\V_J(p)=(p,0,\partial_{u^J}|_p)$ for all $p\in P$. The structure functions of $\T^AP$ are $\varrho^i_\alpha=\rho^i_\alpha$, $\varrho^J_K=\delta^J_K$ and $\mathcal{C}^\alpha_{\beta\gamma}=C^\alpha_{\beta\gamma}$, $\mathcal{C}^\alpha_{\beta K}=\mathcal{C}^J_{\beta K}=\mathcal{C}^J_{K L}=0$.


\subsubsection{The Hamiltonian approach}

Let $\tau:A\rightarrow M$ be a Lie algebroid over a manifold $M$, with anchor $\rho$ and bracket $[\cdot,\cdot]$. As the fibre bundle $P$ we may take $\nu:A^*\rightarrow M$, the dual bundle of $A$. Thus we have defined the $A$-tangent to $A^*$. Taking local coordinates $(x^i)$ on $M$ and choosing a basis $\{e_{\alpha}\}$ of sections of $A$ and the dual basis $\{e^{\alpha}\}$, we have the local coordinates $(x^i,\mu_\alpha)$ on the bundle $A^*$, and we can define the local basis $\{\X_{\alpha},\P^{\alpha}\}$ of sections of $\T^AA^*$ as explained above. We will denote by $\{\X^{\alpha},\P_{\alpha}\}$ the dual basis. We then have,
$
\varrho(\X_{\alpha})=\rho^i_{\alpha}\partial_{x^i}\;\mbox{and}\;\varrho(\P^{\alpha})=\partial_{\mu_{\alpha}},
$
and for a function $f\in\mathcal{C}^\infty(A^*)$ its differential is
$
df=\rho^i_{\alpha}\pd{f}{x^i}\X^{\alpha}+\pd{f}{\mu_{\alpha}}\P_{\alpha}
$.

In the $A$-tangent of $A^*$ there is a canonical section $\theta_A$ of $(\T^A A^*)^*$, called the Liouville section, defined by $\theta_A(a^*)(b,v)=a^*(b)$, for $(b,v)\in \T_{a^*}A^*$, and a canonical symplectic section $\omega_A=-d\theta_A$. In coordinates, they are given by
$$
\theta_A=\mu_{\alpha}\X^{\alpha} \quad\mbox{and}\quad\omega_A=\X^{\alpha}\wedge\P_{\alpha}+\frac{1}{2}C^{\gamma}_{\alpha\beta}\mu_{\gamma}\X^{\alpha}\wedge \X^{\beta}.
$$
The Hamiltonian section $\X_H\in {\rm Sec}(\T^AA^*)$ defined by a function $H\in C^\infty(A^*)$ is written in local coordinates
$$
\X_H=\frac{\partial H}{\partial \mu_{\alpha}}\X_{\alpha}-\left(C^{\gamma}_{\alpha\beta}\mu_{\gamma}\frac{\partial H}{\partial \mu_{\beta}}+\rho_{\alpha}^{\,i}\frac{\partial H}{\partial x^i}\right)\P^{\alpha},
$$
and then,
the VT in the Hamiltonian formalism on a Lie algebroid is
\begin{equation}\label{VT-Hf1}
\DD<\rho^{\,i}_{\alpha}\frac{\partial H}{\partial \mu_{\alpha}}\frac{\partial G}{\partial x^i}-\rho_{\alpha}^{\,i}\frac{\partial H}{\partial x^i}\frac{\partial G}{\partial \mu_{\alpha}}-C^{\gamma}_{\alpha\beta}\mu_{\gamma}\frac{\partial H}{\partial \mu_{\beta}}\frac{\partial G}{\partial \mu_{\alpha}}>=0, \quad  G\in C^\infty(A^*) .
\end{equation}

\begin{example}
Consider a finite-dimensional Lie algebra $\mathfrak{g}$ as a Lie algebroid over a singleton $M=\{e\}$. For a Hamiltonian $H\in C^\infty(\mathfrak{g}^*)$ and a linear virial function $G(\mu)=\langle a, \mu\rangle$, with $a\in\mathfrak{g}$ a constant vector, the VT becomes $\dd<\mathrm{ad}^*_{\pd{H}{\mu}}\mu>=0$. Taking a local basis on $\mathfrak{g}$ and the corresponding linear coordinates on $\mathfrak{g}^*$ we get $\dd<\mu_\gamma C^{\gamma}_{\alpha\beta}\pd{H}{\mu_\beta}a^\alpha>=0$, where $C_{\alpha\beta}^\gamma$ are the structure constants.  Since $a$ is arbitrary we get $\dd<\mu_\gamma C^{\gamma}_{\alpha\beta}\pd{H}{\mu_\beta}>=0$ for every $\alpha=1,\ldots,\dim\mathfrak{g}$.

An important particular case is that of a free rigid body. The Lie algebra is $\mathfrak{g}=\mathfrak{so}(3)$ and the Hamiltonian is defined by $H(\mu)=\frac{1}{2}\mu\cdot I^{-1}\mu$, where $I$ is the inertia tensor. The VT tell us that each component of the cross product $I^{-1}\mu\times\mu$ has vanishing time average.
\end{example}

\subsubsection{The Lagrangian approach}

Similarly, in the general construction above, we can choose $\tau:A\to M$ as the bundle $P$. A regular Lagrangian function $L\in C^\infty(A)$, defines a symplectic section on the $A$-tangent bundle to $A$ as follows. The Cartan 1-section is defined by
$
\theta_L(a,b,v)=\frac{d}{ds}L(a+sb)\Big|_{s=0}$, for $(a,b,v)\in\T^AA$, and then the Cartan 2-section is $\omega_L=-d\theta_L$, which is symplectic provided that the Lagrangian $L$ is regular. The energy $E_L\in C^{\infty}(A)$ associated to $L$ is given by $E_L=\Delta(L)-L$, where $\Delta$ is the Liouville dilation vector field on the vector bundle~$A$.
In local coordinates, the symplectic form $\omega_L$ is given by
$$
\omega_L =\pd{^2L}{y^\alpha\partial y^\beta}\mathcal{X}^\alpha\land \mathcal{V}^\beta
  +\frac{1}{2}\left(\pd{^2L}{x^i\partial y^\alpha}\rho^i_{\,\beta}
  -\pd{^2L}{x^i\partial y^\beta}\rho^i_{\,\alpha}+\pd{L}{y^\gamma}C^\gamma_{\alpha\beta}\right)\mathcal{X}^\alpha\land \mathcal{X}^\beta,
$$
The dynamical section $\Gamma_L$, determined by the equation $i_{\Gamma_L}\omega_L=dE_L$, is $\Gamma_L=y^\alpha\X_\alpha+f^\alpha\V_\alpha$ with $f^\alpha=W^{\alpha\theta}\Bigl(\rho^i_\theta\pd{L}{x^i}-\rho^i_\beta y^\beta\pd{^2L}{x^i\partial y^\theta}-C^\gamma_{\theta\beta}y^\beta\pd{L}{y^\gamma}\Bigr)$,  where $[W^{\alpha\beta}]$ is the inverse matrix of $\bigl[\partial^2L/\partial y^\alpha \partial y^\beta\bigr]$. In the above expressions $\{\X_\alpha,\V_\alpha\}$ denotes a basis constructed as in general case and $\{\X^\alpha,\V^\alpha\}$ denotes its dual basis.

The differential of a function $G\in C^\infty(A)$ is
$
dG=\rho^i_{\,\alpha} \pd{G}{x^i}\X^\alpha+\pd{G}{y^\alpha}\V^\alpha
$ and therefore the virial theorem states that $\dd<\varrho(\Gamma_L)G>=0$, which locally amounts to $\dd<\rho^i_\alpha y^\alpha\pd{G}{x^i}+f^\alpha\pd{G}{y^\alpha}
>=0$, or explicitly
\[
\DD<
\rho^i_\alpha y^\alpha\pd{G}{x^i}+W^{\alpha\theta}\Bigl(\rho^i_\theta\pd{L}{x^i}-\rho^i_\beta y^\beta\pd{^2L}{x^i\partial y^\theta}-C^\gamma_{\theta\beta}y^\beta\pd{L}{y^\gamma}\Bigr)\pd{G}{y^\alpha}
>=0.
\]

\begin{example}
Consider a Lagrangian function $L$ on a finite-dimensional Lie algebra $\mathfrak{g}$, that we consider as a Lie algebroid over a point. For a constant vector $a\in\mathfrak{g}$ we consider the virial function $G(y)=a^\beta\pd{L}{y^\beta}$. The VT becomes $\dd<\mathrm{ad}^*_{y}\pd{L}{y}>=0$, which in quasi-coordinates reads
$
\dd< \pd{L}{y^\gamma}C^{\gamma}_{\alpha\beta}y^\alpha>=0
$,
where we already took into account that $a$ is arbitrary.

In the particular case of a free rigid body, we have $\mathfrak{g}=\mathfrak{so}(3)$ and the Lagrangian is $L(\omega)=\frac{1}{2}\omega\cdot I\omega$, where $I$ are the inertia tensor. It follows that, $\dd<\omega\times I\omega>=0$, in concordance with the result in the Hamiltonian formalism.
\end{example}

Let $\sigma$ be a section of $A$, $\sigma^c$ its complete lift to $\T^AA$, and take as virial function $G=\langle\theta_L,\sigma^c\rangle$. Then, as it was proved in~\cite{EM01} we have $d_\Gamma G=d_{\sigma^c}L$, or in other words $\{G,E_L\}=d_{\sigma^c}L$. Therefore we have proved the following result.

\begin{theorem}
Let $\sigma$ be a section on the Lie algebroid $A$ and let $\sigma^c$ be its complete lift to $\T^AA$. Assume that
 $G=\langle\theta_L,\sigma^c\rangle$ is bounded on its time evolution. Then $\DD<\varrho(\sigma^c)(L)>=0$.
\end{theorem}

\begin{example}
A heavy top can be modeled on the Lie algebroid $S^2\times\mathfrak{so}(3)\to S^2$ with Lagrangian $L=\frac{1}{2}\omega\cdot I\omega - mgl\gamma\cdot e$ (see~\cite{EM01} for the notation and other details). Taking the linear function $G=a\cdot\gamma$ and applying the virial theorem we get $\dd<a\cdot(\gamma\times\omega)>=0$, and since $a$ is arbitrary we arrive to $\dd<\gamma\times\omega>=0$.

On the other hand, we consider a constant vector $a$ on $\mathbb{R}^3\equiv \mathfrak{so}(3)$ and the associated constant section of $A$ given by $\sigma(\gamma)=(\gamma,a)$. The complete lift of $\sigma$ is $\sigma^c=a^i\X_i-(a\times \omega)^i\V_i$. Applying preceding theorem
 we get that $\dd<\rho(\sigma^c)L>=0$, and after an straightforward computation and taking into account that $a$ is arbitrary we arrive at
$
\dd<\omega\times I\omega>=mgl\dd<\gamma\times e>
$.
\end{example}



\end{document}